# Controlled Stability of Molecular Junctions

*D. Dulić, F. Pump, S. Campidelli, P. Lavie, G. Cuniberti, A. Filoramo*

Using single molecules as the building blocks of electronic devices is the ultimate goal of molecular electronics. However, measuring, understanding and manipulating the transport of charge through molecules attached to nanosize electrodes remains a difficult task. It is thus important to fabricate molecular junctions exhibiting reproducible and stable electrical response, and evaluate their performance.[1,2] While chemists are able to synthesize molecules with amazing electronic, optical, magnetic, thermoelectric or electromechanical properties and with the potential to provide electronic devices with novel functionality, the integration into single molecular devices remains the main problem of the field "molecular electronics". It has been demonstrated that the electronic transport through a molecule depends not only on its intrinsic properties but also on specific details of the contacts and the local environment.[3] Here we report a new strategy to monitor the conductance of such molecular junctions enabling to gain further insight into the detailed nature of the conductance of single molecules. Due to the stability and reproducibility of our junctions, our method provides a new perspective on studies of electronic transport on the nanometer scale.

The first consistent results for the conductivity of a molecule in solution were achieved using a statistical analysis of data collected by the Scanning Tunneling Microscope (STM) break junction method.[4] By repeatedly moving the STM tip into and out of contact with the molecular monolayer deposited on the substrate electrode, thousands of molecular junctions can quickly be created and statistical analysis permits to determine the conductance of the molecule with the most probable contact geometries. Another elegant way, the mechanically controllable break junction (MCBJ) technique, has also been proposed to control the spacing between two metallic electrodes with sub-atomic ($< 10^{-10}$ m) resolution.[5] However, the pioneering MCBJ experiments[6-9] did not employ statistical approaches and therefore the number of molecules measured was uncertain.[4,7] Recently, first MCBJ experiments using the statistical approach on alcanedithiol molecules in solution have been reported[10] showing excellent agreement with earlier STM break junction one.[11]

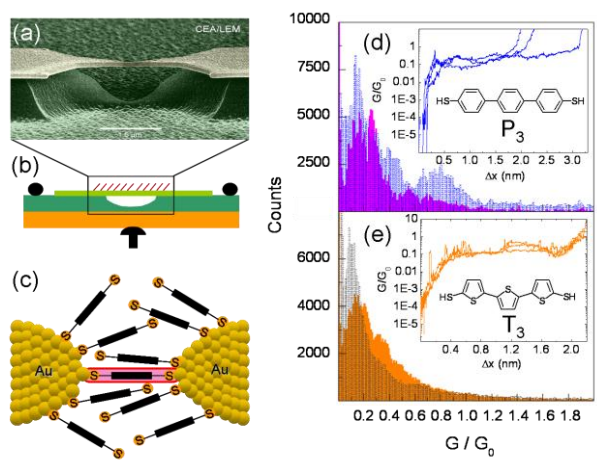

*Figure 1.* a) SEM micrograph of a break junction. b) Layout of the MCBJ technique. c) Schematic vision of the molecule arrangements in the break junction. d) and e): Conductance histograms (repetitional statistics) at V = 20 mV, constructed from 100 traces for two different samples with P3 molecules and two different samples with T3 molecules, respectively; insets show typical conductance traces for P3 and T3.

Here, we study and analyze statistically the electronic transport in terthiophene (T3) and terphenylene (P3) molecules covalently bonded (*via* thiol groups) to two gold electrodes obtained by the MCBJ technique in vacuum (Figure 1). To obtain the conductance for the most probable contact geometry, we first adopted a statistical approach similar to one described in the literature.[10] Details of the experiment are given in the Supporting Information (SI). The P3 and T3 molecules were synthesized according to the methods described in the literature.[12,13] Their structures are shown in Figures 1d and 1e, respectively. These molecules were chosen because of their similar lengths, conjugation paths and coupling groups, while, the presence of additional sulfur atoms in the backbone of T3 modifies the orbital distribution and thus influences the transport characteristics.

For each sample we performed a series of about 100 opening-closing cycles while monitoring the current (for details see SI). From now on we refer to this method as *repetitional statistics*, since each of the individual files used to build the histogram represents a molecular junction with a different conformation with respect to the electrodes. Representative closing traces for P3 and T3 are shown in the insets of Figures 1d and 1e, respectively. The observed conductance plateaus are remarkably long and their lengths are roughly the ones of the molecules, approximately 1.4 nm for P3 and 1.2 nm for T3. In the case of T3 (Figure 1e), the conductance histograms of different MCBJ samples have similar peak values but different peak widths (single Gaussian fit). The P3 histograms (Figure 1d) contain more structure than those of T3. However, as reported in literature, this robust statistical method provides a reliable value for the conductance but at the expense (because of peak widths) of fine-structure in the main peak, which gives information about fluctuations in molecular junctions.[14-16]

Since our MCBJ electrodes can be controlled at sub-atomic resolution, and in the presence of the molecules has impressive stability (on the order of weeks), we developed a new statistical approach referred to hereafter as *fluctuational statistics*, which enables following the time evolution of molecular junctions, fully complementing the repetitional statistics data. As shown below, this method allows mapping the observed conductance peaks corresponding to the fine structure of the main peak. To implement it we blocked the inter-electrodes distance (by stopping the motor) at the position of a conductance plateau and measured the current through the junction as a function of time (left panels of Figure 2, for details see SI).

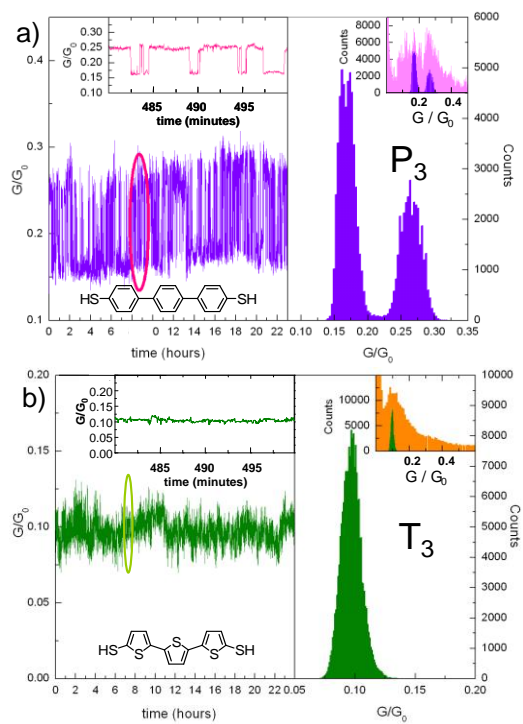

***Figure 2.*** Left panels: Conductance versus time for P3 (a) and T3 (b) molecules. The circled regions are zoomed in the insets. Data points recorded every 0.5 s at 100 mV bias voltage. Right panels: corresponding fluctuational histograms (violet for P3 and green for T3). They are compared in the insets with the repetitional histograms (pink for P3 and orange for T3).

There is one major difference between the P3 and T3 molecular junctions. For P3 the conductance of the molecule changes, typically every few minutes, between two distinct levels (Figure 2a) while not for T3 (Figure 2b). The corresponding conductance histograms are traced in the right panels of Figure 2. Stochastic variations in conductance have previously been observed in thiol-terminated molecules by STM or conductive atomic force microscopy imaging technique.[17-20] They were related to "blinking" of the thiol-gold bond, an intrinsic feature of any molecule connected to a gold surface via thiol groups.[18] However, the T3 molecule does not show such well defined "quantized changes" in conductance.

The conductance switching observed for P3 may be due to a thermally induced change in the bonding sites of the thiol groups terminating the molecules on the electrode surfaces. It is known that the sulfur atoms prefer to bind at high-symmetry positions on the gold surface (top, bridge, hollow-fcc, and hollow-hcp sites assuming a (111) surface), and that the bonding site has an important influence on the conduction properties of single molecules between metallic electrodes.[21,22] Thus, we computed the transmission function through the P3 molecule attached to gold electrodes using the gDFTB method,[23] a combination of the non-equilibrium Green function technique and the density-functional-based tight-binding method DFTB.[24,25]

Comparing the zero-bias conductance values (defined as $G = G_0 T(E_F)$ with $G_0 = 2e^2/h$; for more details and the corresponding transmission curves see SI) obtained for the four different bonding sites we find that the bonding site has a huge influence on the conductance (Table 1).

*Table 1.* Calculated conductance of P3 molecule for different bonding sites between sulfur atoms and Au(111) surface.

| Bonding site | G [ $10^{-2}$ x $G_0$ ] |
|---|---|
| Hollow fcc | 0.18 |
| Hollow hcp | 0.19 |
| Top | 6.16 |
| Bridge | 0.32 |

It is likely that the thermal motion of the connecting sulfur atom on the electrode surface at room temperature can vary the conductance through the molecule, and the stability of the bonding at the high-symmetry points could be the reason for the distinct peaks identified in the statistical analysis in Figure 2a. However, for T3 the presence of a sulfur atom in the ring could increase the overall stability of the bonding,[26] explaining the absence of observable "blinking" in the junction. Note that the fluctuational histograms show improved resolution compared to the repetitional ones, as reported in the right insets of Figure 2.

As shown in Figure 3, the analysis of the conductance on longer time-scales and different voltage reveals further interesting features. The P3 molecule exhibits again a richer evolution of conductance (inset of Figure 3a). In particular, between 110 and 190 hours the conductance oscillates between two distinct levels (independently on the applied voltages). This switching behavior does not appear immediately after formation of the junction, suggesting a local voltage/current activated effect.[27] More generally, all molecular junctions drift on long time-scales due to progressive and gradual rearrangements of the molecules and charge distribution. As discussed below, the comparison between repetitional and fluctuational statistics sheds light on the conductance evolution as the local environment of the connected molecule(s) is modified.

The fluctuational histogram for P3 exhibits distinct peaks at 0.15, 0.20, 0.25, and 0.46 $G_0$ (violet histogram in Figure 3a). These values are consistent with those we obtained from the repetitional statistics (pink histogram in Figure 3a) and correspond to either hopping between hollow-bridge configurations or additional trapping of molecule(s). Differences between the two statistical approaches are more evident on longer time-scales. In particular, the field-induced "wearing on" of the junction by surrounding molecules becomes more important. We emphasize that the time evolution of T3 conductance does not exhibit any switching, even though different bonding site configurations were eventually observed in the repetitional approach.[28] As reported in Figure 3b (inset), upon varying the voltage the conductance versus time changed from 0.1 $G_0$ to 0.2 $G_0$ before increasing further, never returning to the initial value. We interpret this observation by postulating that the near-integer multiples of the fundamental conductance correspond to the formation of a multiple molecular junction where a small number of molecules is connected in parallel between the electrodes. That is, two molecules are bridging the junction.[4] In these data, note an abrupt increase in conductance to 0.2 $G_0$ at 200mV bias voltage. To confirm that increasing the voltage promotes the collection of another molecule we formed a fresh molecular junction and repeated the experiment at 100mV (see SI).

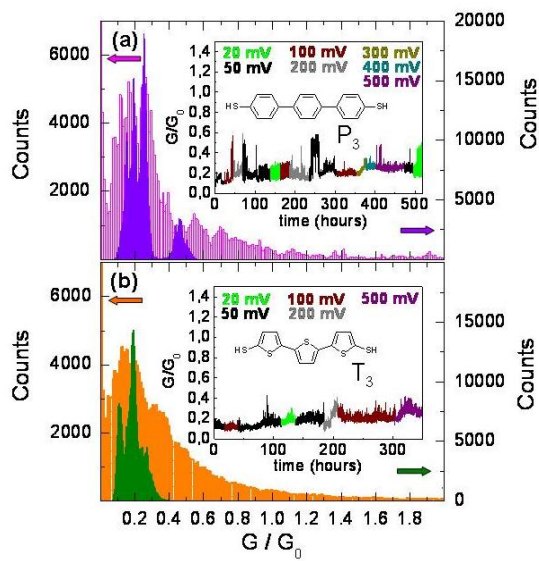

*Figure 3.* a) and b): Conductance versus time (insets) with changing bias voltage and comparison (main figures) between repetitional (pink and orange) and fluctuational (violet and green) statistics for P3 and T3 respectively. Points were taken every 0.5 s. For P3, stochastic switching is observed between 110 and 190 hours.

Our conductance values are substantially higher than those obtained by the repetitional statistics STM break junction method in liquid for similar molecules.[14] Theoretical calculations[22] show that the conductivity of a molecular junction is extremely sensitive to the geometry of the contact, which in our case is different from the STM geometry, and the presence of different media might play a role. Moreover, we do not claim that the most probable conductance value obtained by the repetitional statistics method corresponds solely to the conductance of a single molecule or exclude the possibility that the overall conductance is influenced by the presence of neighboring molecules. However, our data show that the junction persists for a long time only when the bridging molecule(s) is (are) connected to both electrodes, and that the conductance obtained by fluctuational statistics corresponds to the most probable value obtained by the repetitional statistics method. Finally, note that the stochastic changes in STM imaging experiments [19] do not occur in experiments with larger number of the same molecules.[29,30]

In conclusion, we have studied the influence of the interface on simple molecular wires, a factor essential to understand before molecular electronic devices can be realized. Due to the week-long stability and reproducibility of our junctions, our method opens new horizons in studying molecules with more interesting intrinsic properties, as well as designing and studying molecules with different end groups that could suppress stochastic switching. This will have an impact not only on the field of single-molecule electronics, but also in general molecular electronics dealing with self-assembled monolayers.

**Keywords:** molecular electronics · break junction · electron transport · sulfur-gold interactions · single molecule studies